\documentstyle[12pt]{article}
\newcommand{\be}{\begin{equation}}
\newcommand{\ee}{\end{equation}}

\newcounter{tempfigc}			
\newcommand{\fcaption}[1]{
	\addtocounter{tempfigc}{1}
{\noindent\parbox{16.5 cm} {\tenrm Fig.~\thetempfigc. #1} }}

\newcounter{temptabc}
\newcommand{\tcaption}[1]{			
	\addtocounter{temptabc}{1}
{\noindent\parbox{16.5 cm} {\tenrm Tab.~\thetemptabc. #1} }}

\begin{document}

\thispagestyle{empty}
\begin{titlepage}

\begin{flushright}
\begin{minipage}[t]{4cm}
\begin{flushleft}
{\baselineskip = 14pt
hep-ph/9507360\\
July, 1995\\
}
\end{flushleft}
\end{minipage}
\end{flushright}

\vspace{0.5cm}

\begin{center}
\Large\bf
The electro-magnetic Form Factors of the Proton in chiral Soliton Models
\footnote{Contribution to the Sixth International Symposium on
Meson-Nucleon Physics\\ and the Structure of the Nucleon,
Blaubeuren/Tuebingen, Germany, 10-14 July, 1995}
\end{center}

\vspace{0.5cm}
\vfill

\begin{center}
{\large
Gottfried {\sc Holzwarth}\footnote{
e-mail address : {\tt holzwarth@hrz.physik.uni-siegen.d400.de}}
}
\\
{\it Fachbereich Physik, Universitaet Siegen,
57068 Siegen, Germany}
\end{center}

\vfill

\vspace{0.5cm}

\begin{abstract}
\baselineskip = 15pt

The electro-magnetic form factors of the proton are calculated in a
chiral soliton model with relativistic corrections.
The magnetic form factor $G_M$ is shown to agree well with the
new SLAC data for spacelike $Q^2$ up to 30 (GeV/c)$^2$ if
superconvergence is imposed. The direct
continuation through a Laurent series to the timelike region above the
physical threshold is in fair agreement with the presently available
set of data.
The electric form factor $G_E$ is dominated by a zero in the few
(GeV/c)$^2$ region which appears to be in conflict with the SLAC data.
\end{abstract}
\end{titlepage}

\setcounter{footnote}{0}

\section{Relativistic soliton form factors}

The new SLAC data$^{1,2}$ for electro-magnetic form factors (FF)
of the proton to high $Q^2$ pose a challenging test for
the relativistically corrected FFs of chiral soliton models.

 It has repeatedly been demonstrated for
various versions of chiral lagrangians that the nucleon e.m. FFs
are rather well accounted for low $Q^2$ with nucleons as nonrelativistic
solitons in coupled $\pi, \varrho$, and $\omega$ fields$^{3,4}$.
The implementation of relativistic corrections
is especially easy for solitonic nucleons
due to the Lorentz covariance of the field equations (in contrast to
the corresponding problem in bag models$^5$).
 The corrections reflect the Lorentz
boost from the soliton rest frame to the Breit frame, in which the
soliton moves with velocity $v$ which satisfies
\be
\gamma^2=(1-v^2)^{-1} = 1 + \frac{Q^2}{(2M)^2}
\ee
for momentum transfer $Q^2$ ($Q^2>0$ in the spacelike region) and
soliton mass $M$.
The classical result for the magnetic FF is$^6$
\be
G_M (Q^2) = \frac{1}{1 + \frac{Q^2}{(2M)^2}} G^{nr}_M \left(
\frac{Q^2}{1 + \frac{Q^2}{(2M)^2}} \right) \; ,
\ee
where $G^{nr}$ is the nonrelativistic FF evaluated in the
soliton restframe.
The electric FF $G_E$
does not contain the factor $\gamma^{-2}$ on the right-hand side$^6$:
\be
G_E (Q^2) = G^{nr}_E \left(
\frac{Q^2}{1 + \frac{Q^2}{(2M)^2}} \right) \;
\ee
 (this is in contrast to bag models$^5$ where the wave functions of
the spectator quarks supply the factor $\gamma^{-2}$ also for $G_E$.)

According to the derivation of (2,3) within the tree approximation of
the soliton model $M$ is the classical soliton mass $M_S$, although
ideally, of course, $M$ should coincide with
the physical nucleon mass $M_N$.
{}From (2,3) the asymptotic limit of $G(Q^2)$ for $Q^2 \rightarrow
\infty$ is given by $G^{nr}(4M^2)$.
For commonly used chiral lagrangians the first zeros of the
nonrelativistic FFs occur at masses $M_0$
\be
G^{nr}(4 M^2_{0}) = 0
\ee
which are of the order of the nucleon mass, with $M_0<M_N$ for
$G_E^{nr}$ and $M_0>M_N$ for $G_M^{nr}$. This implies that the
asymptotic behaviour of $Q^4G(Q^2)$ is very sensitive to the precise
value of $M$ used in (2,3):
\be
\lim_{Q^2 \rightarrow \infty} Q^4G(Q^2) = \pm \infty
  \mbox{~~~~for~~~~}  M \begin{array}{c} <\vspace{-3mm}\\>\end{array} M_0.
\ee
The actual values of $M_0$ for which $G^{nr}(4M_0^2)$ vanishes, depend
on the choice of the parameters in the effective lagrangian;
furthermore both, $M_S$ and $G^{nr}$ are subject to quantum
corrections. It is therefore unrealistic to expect reliable predictions
from the model itself for the high-$Q^2$ behaviour of $Q^4G(Q^2)$.

This ambiguity in  the high-$Q^2$ behaviour of $Q^4G(Q^2)$ can be used
to {\it {impose}} superconvergence $( Q^2 G_M (Q^2) \to 0 \quad {\mbox{{for}}}
\quad Q^2
\to \infty )$ on $G_M(Q^2)$ by choosing $M=M_0$ in (2), or, to
put it more generally, to check the functional form of (2)
against the experimentally observed behaviour of $Q^4G(Q^2)$ for large $Q^2$
by choosing $M$ as an adjustable parameter.
Due to the lack of the factor $\gamma^{-2}$ on the right
hand side in (3), superconvergence cannot be imposed on $G_E$
by any choice of $M$. For a specific effective lagrangian (and due to
possibly different quantum corrections) we also should not expect $M$
to be necessarily the same for different formfactors.

The low-$Q^2$
behaviour is not strongly affected by these variations in $M$,
although due to the factor $\gamma^{-2}$ in front of $G_M^{nr}$ in (2),
even the magnetic radius receives a small contribution from finite
values of $M$.

\section{The minimal $\pi$-$\rho$-$\omega$ model}

In order to study the implications of a simple effective lagrangian
we choose the
minimal model which comprises $\rho$ and $\omega$  mesons
together with the pionic field $U$ in chiral covariant way:
\be
{\cal L}_{V\!M}={\cal L}^{(2)}+{\cal L}^{(\rho)}+{\cal L}^{(\omega)}
\ee
with
\be
{\cal L}^{(2)}=\frac{f_\pi^2}{4}\int (-trL_\mu L^\mu+m_\pi^2
tr(U+U^\dagger-2))d^3x,
\ee
\be
{\cal L}^{(\rho)}= \int \left(-\frac{1}{8} tr \rho_{\mu\nu} \rho^{\mu\nu}
+\frac{m_\rho^2}{4} tr(\rho_\mu
-\frac{i}{2g}(l_\mu-r_\mu))^2 \right) d^3x,
\ee
\be
{\cal L}^{(\omega)}=\int \left(-\frac{1}{4} \omega_{\mu\nu}
\omega^{\mu\nu} +\frac{m_\omega^2}{2}
\omega_\mu \omega^\mu +3g_\omega \omega_\mu B^\mu \right) d^3x,
\ee
the Maurer-Cartan forms
\be
L^\mu = U^\dagger \partial^\mu U = L^\mu_a \tau_a,
\ee
topological baryon current $B_\mu$
\be
B_\mu=\frac{1}{24 \pi^2} \epsilon_{\mu\nu\rho\sigma} tr L^\nu L^\rho
L^\sigma
\ee
and $l_\mu=\xi^\dagger \partial_\mu \xi, \;r_\mu=\partial_\mu \xi
\xi^\dagger$
with $\xi=U^{\frac{1}{2}}$.
In the gauge transformation of the vector mesons $V_\mu$
\be
V^\mu \rightarrow  e^{(i \epsilon_0 Q_0+i \epsilon_V Q_V)}
(V^\mu +\frac{Q_V}{g} \partial^\mu \epsilon_V
+\frac{Q_0}{g_0} \partial ^\mu \epsilon_0)
e^{(-i \epsilon_0 Q_0-i \epsilon_V Q_V)}
\ee
(with $Q_0=1/6 \ , \ Q_V=\tau_3 /2$)
through which the electromagnetic currents are defined,
the gauge parameter $g_0$ need not coincide with $g_\omega$
because the contribution of the neutral $\omega$-mesons to the
isoscalar part of the e.m. current is not necessarily fixed through
the electric charge e(=1).

With the experimental values for $f_\pi$, the meson masses
$m_\pi$, $m_\rho$, $m_\omega$, and $g$ fixed by the KSRF relation
$g=m_\rho/(2\sqrt{2} f_\pi)$ = 2.925,
${\cal L}_{V\!M}$ contains $g_\omega$ as the only free parameter;
we use it to fit the magnetic moment of the proton
$\mu_p=G_M^p(0)=2.79$; the resulting value is $g_\omega=4.125$.

\section{Results}

For this choice of the effective ${\cal L}_{V\!M}$ the low-$Q^2$
pattern of the FFs is still sensitive to the value of $g_0$
in the isoscalar part of the e.m. current.
Agreement with the data for $G_E$ in the region $Q^2<1$ (GeV/c)$^2$
can be achieved for $g_0 \geq 2.5 g_\omega$.
For $g_0 = 2.5 g_\omega$ superconvergence
for $G_M$ requires $M=1.12$ GeV in (2).
The resulting e.m. FFs for the proton (divided through the standard
dipole $G_D=(1+Q^2/0.71)^{-2}$) are shown by the full lines
in figs. 1 and 2, plotted against the logarithm of $Q^2$.
Both, $G_E$ and $G_M/\mu_p$, are calculated for the {\it same} value
of $M$(=1.12 GeV).
The rapid decrease of $G_E/G_D$ above $Q^2 \sim 1$ (GeV/c)$^2$
which is in apparent contradiction to the SLAC data,
has its origin in the first zero of $G_E^{nr}$ which is pushed up to
$Q^2\approx $ 3.7 (GeV/c)$^2$ by the boost to the Breit frame in (3).
It can be shifted to higher $Q^2$ by decreasing $M$ but then
$G_E$ overshoots the dipole near $Q^2\approx 1$ (GeV/c)$^2$
(the dash-dotted line in fig.2 is calculated for $M=0.94$ GeV).
Because the rapid decrease of $G_E/G_D$ is due to a zero in $G_E^{nr}$
it cannot be removed by an additional factor
$\gamma^{-2}$ in front of $G_E^{nr}$ which may appear in bag models.

For the high-$Q^2$ part of $G_M$ the choice $g_0=g_\omega$ seems
preferable, which then requires $M=1.13$ GeV for superconvergence
(dashed line in fig.1). However, this impairs the quality of
agreement at low $Q^2$ for $G_E$. Only with a value
of $M$ smaller than the nucleon mass  the zero in
$G_E$ can be pushed up to about 10 (GeV/c)$^2$ so that the SLAC
data can be accommodated. But then the overall pattern of $G_E$
is very unsatisfactory (the dashed line in fig.2 is calculated with
$M=0.76$ GeV).

Although details depend on the choice of parameters
in the effective lagrangian and in the isoscalar part of the e.m.
current
it is evident from figs.1 and 2 that the functional form (2)
is able to describe the general pattern of the observed magnetic FF
over the whole range of measured $Q^2$ values if superconvergence
is imposed, without any further "QCD" corrections$^8$. The electric FF
is dominated by a zero in the few (GeV/c)$^2$ region which is very
difficult to avoid and appears to be in conflict with the SLAC data.
For $g_0 = 2.5 g_\omega$ it is possible to satisfy the scaling relation
$G_M/\mu_p =G_E$
with very good accuracy up to $Q^2\approx 1$ (GeV/c)$^2$ which is
quite remarkable for a model in which the Besselfunctions $j_0$ and
$j_1$ determine the electric and magnetic FFs, respectively, (which
naively implies for the ratio of the radii $<r^2_M>/<r^2_E>\sim 3/5$).
Clearly, more experimental information on the proton electric FF
in the few (GeV/c)$^2$ region would be very helpful for a critical
assessment of these implications of the soliton model.

\section{Extension to large timelike $Q^2$}

We have seen that, with superconvergence imposed,
the expression (2) reproduces
the essential features of $G_M(Q^2)$ up to the highest
measured values of spacelike $Q^2$.
 A peculiar consequence of (2,3) is
that the argument of $G^{nr}$ in (2,3) is positive for $Q^2 < -
(2M)^2$, i.e. $G(Q^2)$ is real for timelike $Q^2$ beyond the $N\bar N$
threshold. This may indicate an unphysical feature of (2,3) which maps
$G^{nr}(q^2)$ for large
$q^2$ onto $G(Q^2)$ with timelike $Q^2$ just beyond the $N\bar N$
threshold. (It should also be noted that $G_M(Q^2)$ does not have a pole at
$Q^2 = -(2 M)^2$, because the divergent factor in front of $G^{nr}_M$ in
(2) is compensated by the vanishing of $G^{nr}_M(q^2)$ for $q^2 \to
\infty$.)
But as a speculation, it is tempting to accept the transformation (2)
 also for large timelike $Q^2$ (corresponding to $q^2 > 4 M^2$ in
$G^{nr}_M(q^2)$) as a prediction for (at least the real part of)
$G_M(Q^2)$. The connection between large space- and timelike values of
$Q^2$ then may be established through a Laurent
expansion of $G_M(Q^2)$ for $|Q^2|\to \infty$:
\be
G_M(Q^2) = \frac{1}{\pi} \int^{(2M)^2}_{t_0} \frac{\Gamma (t')}{t'+Q^2} dt' =
\sum^\infty_{i=0} M^{(i)} \cdot (Q^2)^{-1-i}
\ee
with moments $M^{(i)}$ of the spectral function
\be
M^{(i)} = \frac{1}{\pi} \int^{(2M)^2}_{t_0} \Gamma(t') (-t')^i dt'.
\ee
The continuation to the timelike region beyond the $N\bar N$ threshold
then is simply a matter of changing the sign of $Q^2$ in the Laurent
series (13). With a sufficiently accurate set of data such an analysis
could be done in a model independent way.

Table 1 shows two fits A and B of the series (13) to the function  $Q^4
G_M(Q^2)/\mu_p$ for large spacelike $Q^2$, with 5 and 7 moments,
respectively.  \\

\tcaption{ The moments $M^{(i)}/\mu_p$ in units of $[$GeV$^2]^{1+i}$ as
obtained from two fits of the Laurent series (13) to
$Q^4 G_M^p(Q^2)$ (for $g_0=g_\omega$, i.e. corresponding to
the dashed line in fig.1) for large spacelike $Q^2$ with  5 (7)
nonvanishing moments in fit A(B).}

\hspace*{\fill}\begin{tabular}{|c|c|c|c|c|c|c|c|} \hline
 i = & 1 & $\quad$ 2 $\quad$  & 3 & $\quad$ 4 $\quad$ & $\quad$ 5
$\quad$ & $\; \quad$ 6 $\; \quad$ & $\! \quad$  7 $\! \quad$ \\ \hline
A &  0.2937 &  1.91 & - 12.06 &  28.4 & - 25 & 0 & 0 \\ \hline
B &  0.2937 &  1.91 & - 12.06 &  27.6 & - 22 &  55 &
- 190 \\ \hline
\end{tabular} \hspace*{\fill} \\ [.3cm]

The formfactors $G^p_M$ resulting from the series (13) with the
moments of table 1, for timelike $-Q^2 > 3.5 $ GeV$^2$ are plotted in fig.3.
($G_M$ is negative in this region, fig.3 shows
$|G_M|$ together with the present worldwide set of data for this
quantity$^9$).
The fact that the experimental data for $|G|$ show a slower
decrease may be an indication of the imaginary part missing in the
expression (2) for $G_M(Q^2)$. It appears that $|G|$ is not affected
by the higher moments above $-Q^2 > 5$ (GeV/c)$^2$,
and it is not very sensitive
in the region from 3.5 to 5 (GeV/c)$^2$ as long as
we exclude the possibility of extremely large higher
moments in (13), or strong singularities close to the physical
threshold $-Q^2=4M^2$. In this respect it is interesting
that this continuation of (13) to timelike $Q^2$
reproduces at least the order of magnitude of the form factor
above the physical threshold.

\newpage
{\bf References}

\begin{enumerate}

\item
A.F. Sill et al., Phys.Rev. {\bf D 48} (1993) 29.

\item
L. Andivahis et al., Phys.Rev. {\bf D 50} (1994) 5491;\\
R.C. Walker et al., Phys.Rev. {\bf D 49} (1994) 5671.

\item
U.G. Meissner, N. Kaiser and W. Weise,  Nucl.Phys. {\bf A 466} (1987)
685;\\
N. Kaiser, U. Vogl, W. Weise and U.G. Meissner, Nucl.Phys. {\bf A
484} (1988) 593.

\item
F. Meier, in: Baryons as Skyrme solitons,  ed.
G. Holzwarth (World Scientific, Singapore 1993) p.159.

\item
M.V.Barnhill,  Phys.Rev. {\bf D 20} (1979) 723;\\
M. Betz and R. Goldflam,  Phys.Rev. {\bf D 28} (1983) 2848;\\
X.M. Wang and P.C. Yin,  Phys.Lett. {\bf B 140} (1984)
249;\\
C.J. Benesh and G.A. Miller,  Phys.Rev. {\bf D 36} (1987)
1344.

\item
Xiangdong Ji, Phys.Lett. {\bf B 254} (1991) 456 .

\item
G. H\"ohler et al., Nucl.Phys. {\bf B 114} (1976) 505.

\item
See refs.$^{4-8,30-51}$ in A.F. Sill et al.$^1$.

\item
A. Antonelli et al., Phys.Lett. {\bf B 334} (1994) 431;\\
G. Bardin et al., Nucl.Phys. {\bf B 411} (1994) 3;\\
and references therein.

\end{enumerate}

\newpage

\fcaption{ The magnetic form factor of the proton, $G^p_M/\mu_p$
(divided through the standard dipole $G_D=(1 + Q^2/0.71)^{-2}$)
plotted against the logarithm of the spacelike momentum transfer $Q^2$
as obtained from the model defined in section 2.
Full line: $g_0=2.5 g_\omega$ with $M=1.12$ GeV; dashed line:
$g_0= g_\omega$ with $M=1.13$ GeV;
dots and triangles denote the SLAC data of ref.$^1$ and refs.$^2$,
respectively; open circles show the data compilation of ref.$^7$.}

\vspace{0.5cm}
\fcaption{ The electric form factor of the proton, $G^p_E/G_D$
plotted against the logarithm of the spacelike momentum transfer $Q^2$
as obtained from the model defined in section 2.
Full line: $g_0=2.5 g_\omega$ with $M=1.12$ GeV, (dash-dotted line:
$M=0.94$ GeV);
dashed line: $g_0= g_\omega$ with $M=0.76$ GeV;
triangles denote the SLAC data of refs.$^2$; open circles show the
data compilation of ref.$^7$.  }

\vspace{0.5cm}
\fcaption{ The Magnetic formfactor of the proton $|G|$ for timelike momentum
transfer $t=-Q^2$ above the $N\bar N$ threshold. The dotted (dashed) curve
is the series (13) with the 5 (7) moments of fit A (B) given in table 1.
The dots with error bars show the present
worldwide data set for $|G|$ from ref.$^9$. }

\end{document}